\documentclass[manuscript]{aastex}

\usepackage{graphicx}
\usepackage{epstopdf}

\shorttitle{Plasm Outflows and the Associated Magnetic field}
\shortauthors{Liu et al.}

\begin{document}

\title{Multi-channel Observations of Plasma Outflows and the Associated Small-Scale Magnetic Field Cacellations on the Edges of an Active Region}

\author{S. Liu$^{1}$ and J.T. Su$^{1,2}$}
\affil{$^{1}$National Astronomical Observatory and key Laboratory of Solar Activity, \\Chinese Academy of
Sciences, Beijing 100012, China\\
$^{2}$State Key Laboratory of Space Weather, Chinese Academy of Sciences, Beijing 100190
}



\begin{abstract}
With the SDO/AIA instrument, continuous and intermittent plasma outflows
are observed on the boundaries of an active region along two distinct open coronal loops. 
By investigating the temporal sequence magnetograms obtained from HMI/SDO, it is found that small-scale magnetic reconnection probably plays
an important role in the generation of the plasma outflows in the coronal loops.
It is found that the origin of the plasma outflows coincides with the locations of the small-scale magnetic fields with mixed polarities,
which suggests that the plasma outflows along coronal loops probably results from the magnetic
reconnection between the small-scale close emerging loops and the large-scale open active region coronal loops.
\end{abstract}


\section{INTRODUCTION}
There are large amounts of highly dynamic phenomena with various scales
in the solar atmosphere, which are indicative of strong radiance increasing or
proper motions of bright features (dark when seen in absorption).
Plasma outflows  are usually observed during those highly dynamic events.
It is believed that they may contribute to both the coronal heating and
solar wind generation \citep{mad08}. It is conventionally recognized
that most of the plasma outflows are corresponding to both fast and slow solar winds.
At least, these outflows may be one of the solar wind sources on the Sun \citep{sak07}.
For example, fast solar wind probably emanates from coronal holes \citep{kri73, tu05}.
Slow solar wind probably originates from the boundaries of coronal holes \citep{wan90, koh06, tia11},
helmet streamers \citep{wan90, she97, kil09}, the edges of active regions
\citep{sak07, har08, mar08} and the open field regions in the quiet Sun \citep{he07, tia11}.
Due to the important roles of these plasma outflows in solar wind,
the causes of these outflows with the velocity of tens to hundreds km s$^{-1}$
are worthy to be investigated.

Since the launch of advanced space instruments such as Solar and
Heliospheric Observatory (SoHO), Transition Region and Coronal Explorer (TRACE)
and H\emph{inode}, the outflows along coronal loops/structures emanating from
the boundary of active region have been observed and investigated more frequently.
Using Extreme Ultraviolet (EUV) spectra obtained with the Coronal Diagnostic Spectrometer (CDS) on SoHO,
\citet{bre97} presented significant plasma outflows along coronal and transition active region loops, with velocities of ~50 km/s, or more.
\citet{win01} detected evident outflows along a bundle of coronal active region loops, with velocity ranging from 5 to 20 km/s.
 There is no obvious periodicity in this observation and
they concluded that the outflows are a class of reconnection events, and postulated
these mass flows are probably driven by small scale magnetic reconnection events occurring at the
foot points of coronal loops. With the observations taken by the X-Ray Telescope (XRT) onboard the H\emph{inode},
the outflows originating from the edge of an active region with a speed of $\sim$140 km s$^{-1}$
are identified by \citet{sak07}, and they concluded that these outflows are probably associated with open magnetic field lines
and may contribute to the generation of slow solar wind.
It is also should noticed that there are large areas of plasma outflows at
the boundaries of active region revealed by the Extreme Ultraviolet Imaging Spectrometer (EIS)
on the H\emph{inode} spacecraft, and they were thought to be a possible candidate for the solar wind
 \citep{sak07, doc07, har08}. \citet{bak09} suggested that outflows
may originate from specific locations of the magnetic field topology where field
lines display strong gradients of magnetic connectivity, namely, quasi-separatrix
layer (QSL). With the above observations, the further characterization
of outflows regions should be important for us to understand the fundamental physical processes
that cause these outflows and then the production of the solar wind and mass flows into the corona.
From the above works, we recognize that the magnetic reconnection probably plays a key role
in the generation of plasm outflows. Hence, the associations between plasma outflows and magnetic field evolutions
should deserve to be further investigated.

In this paper, we study some events of plasma outflows in the coronal loops  emanating from the boundaries
of an active region, and find that the cancellation of small scale mixed magnetic fields
at the boundary of the main active region causes the occurrence of the plasma outflows. The rest of this paper is organized as
follows: the description of observations and data reduction and the results
are arranged in Section~\ref{S-Obser and Result}.
Section~\ref{S-Conl} gives the discussions and conclusions.

\section{OBSERVATIONS AND RESULTS}
\label{S-Obser and Result}
Atmospheric Imaging Assembly \citep[AIA;] []{lem12} and the
Helioseismic and Magnetic Imager (HMI) \citep{sch11} are two main instruments onboard the Solar Dynamics Observatory (SDO).
AIA takes full-disk images of the sun in 3 UV-visible and 7 EUV channels with a resolution of 0$\arcsec$.6 pixel$^{-1}$.
HMI obtains full-disk magnetograms
in the photospheric absorption line \ion{Fe}{1}
centered at the wavelength 6173.3 \AA{} with high spatial and temporal
resolutions of 0$\arcsec$.6 pixel$^{-1}$ and 45 s, respectively.
In this study, the channels of 211 \AA{}, 193 \AA{}, 171 \AA{},
304 \AA{}, 1600 \AA{} and 1700 \AA{} are used to detect the plasm outflows
along the coronal loops rooted at the edge of AR NOAA 11504,
which were observed about from 18:00 UT to 23:00 UT on June 14 2012.
It means the data sets cover the whole solar atmosphere from photosphere,
chromosphere, transition regions to corona. We use the product of vector
magnetograms projected and remapped to heliographic coordinates. The data
processing is based on the standard Solar Softwares of these
instruments.

\begin{figure}
\epsscale{1.} \plotone{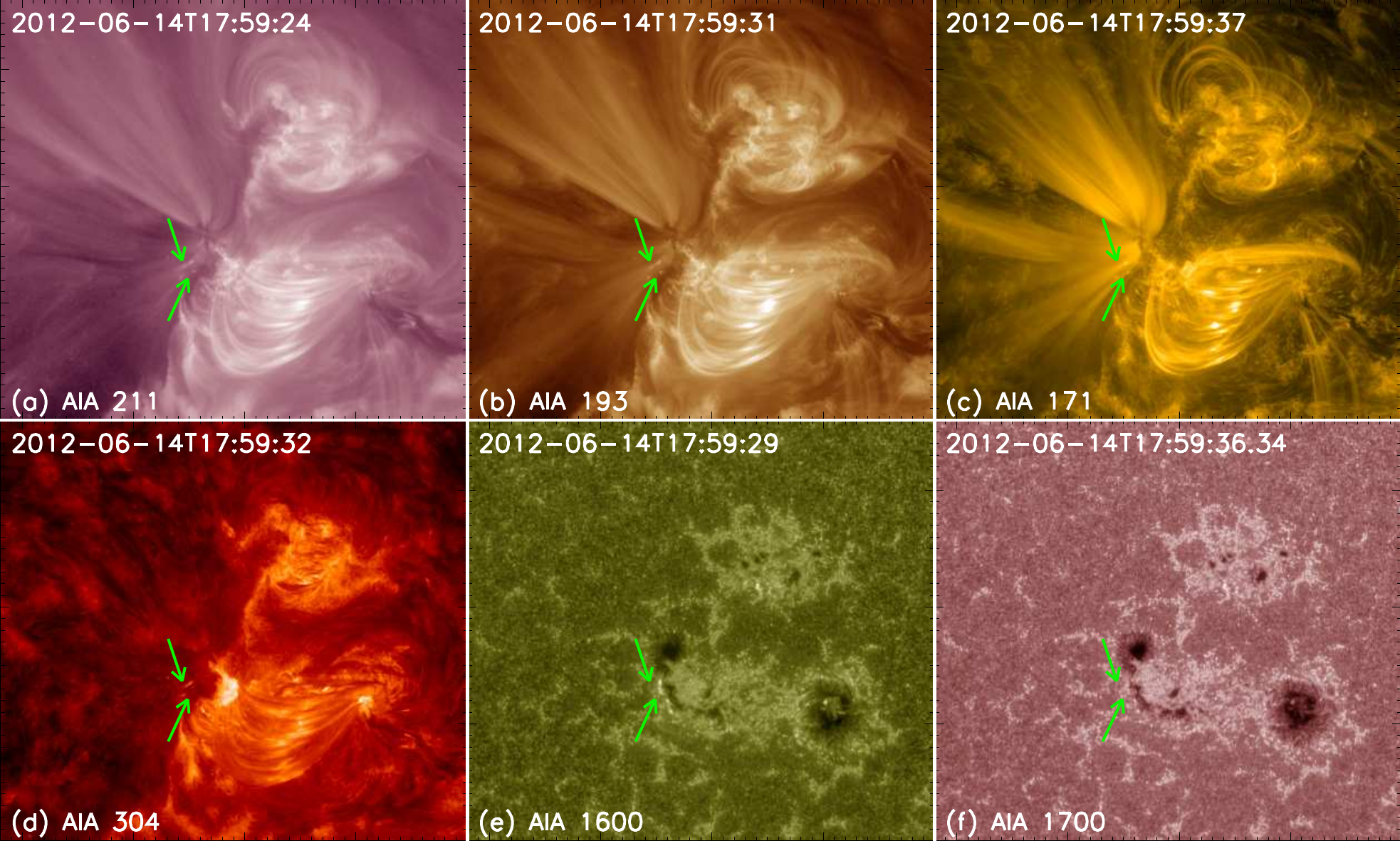}
\caption{Maps of NOAA AR 11504 with AIA observations, where the corresponding channel and observational
time are labeled on each frame. Red (blue) contours represent positive (negative) values of
the superimposed HMI LOS magnetic field.}
\label{Fig1}
\end{figure}

In Fig~\ref{Fig1}, images in the AIA channels of 211 \AA{}, 193 \AA{}, 171 \AA{},
304 \AA{}, 1600 \AA{} and 1700 \AA{} observed at about 18:00 UT on June 14 2012 are
shown, respectively. The corresponding wavelength is plotted on each frame. The plasma outflows can be seen clearly in 211 \AA{},
193 \AA{}, 171 \AA{} and 304 \AA{} channels with the temporal sequence data (see one
movie in 171 \AA{} channel). The locations of coronal loops with outflows
are shown by two green arrows in Fig~\ref{Fig1}. There are obvious bright dots
or intensity enhancements along coronal loops during this observed time interval,
and this phenomenon occurs intermittently. Hence, these plasma outflows of interest
are to be studied in present work.

In Figure 2, panels (a) and (b) correspond to the images of 171 \AA{},
and one light of sight (LOS) magnetogram, respectively. Two virtual slits along
coronal loops with evident outflows are highlighted by two white lines (S1 and S2)
for further investigations. It can been seen that the coronal loops are rooted at the southeast edges of
the main active region, where small-scale magnetic fields with positive polarity, adjacent to
large-scale negative magnetic polarity of the main active region, are concentrated on. Hence
the interactions between the two magnetic systems with opposite polarities probably
play an important role in the production of plasma outflows observed.

\begin{figure}
\epsscale{1.} \plotone{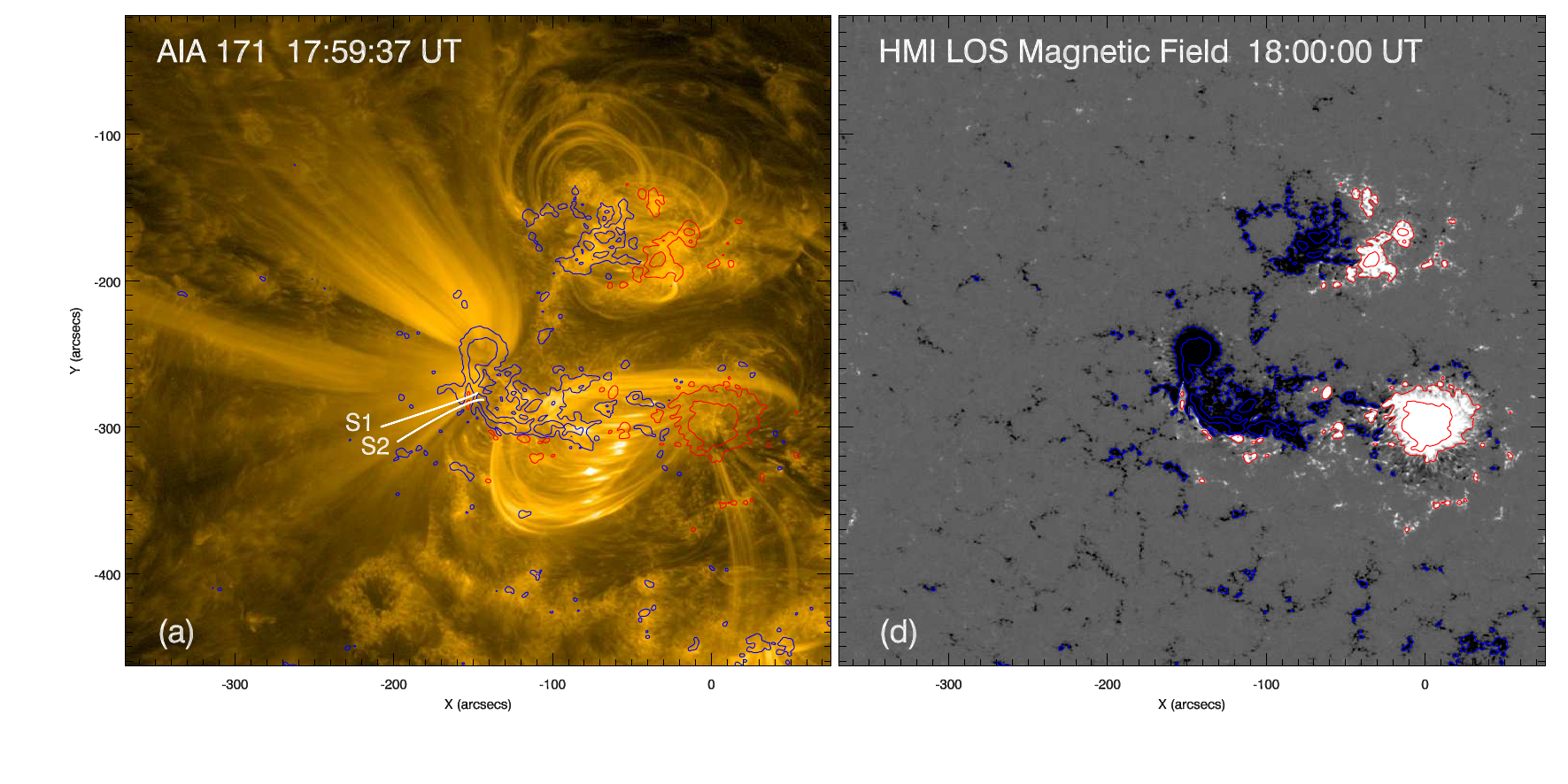} \caption{AIA and HMI observations of NOAA AR 11504,
images of (a) and (b) correspond to 171  \AA{} and LOS magnetogram
labeled. Where white lines plotted in (a) indicate two coronal loops with
evident plasma outflows, and they are labeled by S1 and S2, respectively.
Red (blue) contours represent positive (negative) values of the superimposed HMI LOS magnetic field.}
\label{Fig2}
\end{figure}

\begin{figure}
\epsscale{1.} \plotone{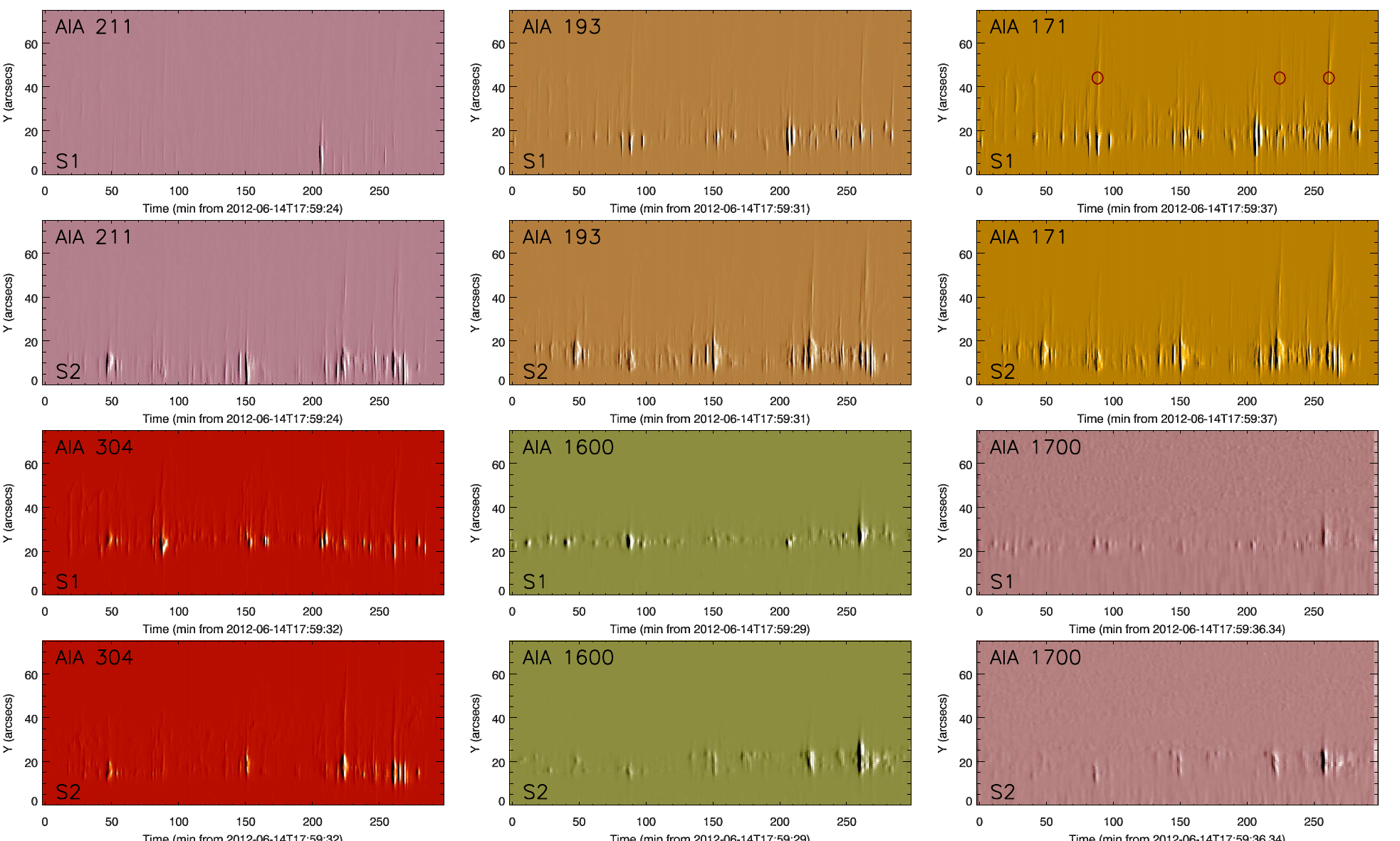} \caption{Time-distance diagrams of
the EUV and UV-visible intensities along the white lines in Figure 2(a).
Rows 1 and 3 correspond slit S1, while Rows 2 and 4 correspond slit S2, respectively.}
\label{Fig3}
\end{figure}

Time-distance diagrams of multi-channel observations for S1 and S2 slits
are presented in Fig \ref{Fig3}, where x-axis indicates the time spanning from 18:00 UT to 23:00 UT, and
y-axis means the distance along the marked coronal loops in Fig \ref{Fig2}. The slope
of bright strips in the diagram means there are plasma outflows propagating along coronal loops.
We can see that the signatures of outflows are most strong in the EUV line of 171 \AA{}, and relatively weak
in the other EUV lines (of 211 \AA{}, 193 \AA{} and 304 \AA{}). Only a few strong outflows can be detected in
the UV lines of 1600 \AA{} and 1700 \AA{}. Therefore, the plasma outflows identified are mutli-thermal.

From top to bottom rows, Fig~\ref{Fig4} shows three typical outflow events in AIA 211 \AA{} channel
(highlighted by three red circles drawn in Fig \ref{Fig3}). The red/blue contours superimposed on each panel
are positive/negative HMI LOS magnetic field. It can be found that those positions with obvious intensity
enhancement (conspicuous bright dots) are exactly located in the regions of magnetic neutral line between
the small-scale positive magnetic fields and the large-scale negative magnetic fields.
Similarly, these outflow events in the others channels of 193 \AA{}, 171 \AA{} and 304 \AA{}, are shown in
Figs~\ref{Fig5}-\ref{Fig7}, respectively.
Their counterparts in 1600 \AA{} and 1700 \AA{} channels are shown in Figs 8 and 9. However, the signatures
of plasma propagating in the UV channels, especially in 1700 \AA{}, nearly vanish. On the whole, the observations
reveal that these plasma outflows can be easily detected at the upper solar atmosphere, and they are weak at the
lower atmosphere.

The evolutions of small-scale positive fields, from which the outflows are originated, are shown in Fig~\ref{Fig10}(a).
The evolution process can be described as follows: Firstly, two white patches with stronger field strength (denoted by
the red contours) appears at 17:59 UT, then they approach each other likely squeezed by the two black patches with negative
polarity from two opposite directions (shown by the two arrows). Thus, the two white patches begin to combine together.
Subsequently, at 19:59 UT a small white patch begins to separate from the whole one. In the panels, three dot lines
delineating movement of the three patches with positive polarity are plotted, which can be used to estimate combining
and separating speeds of the white patches. We obtain the combine speed in magnitude $\sim$0.5 km s$^{-1}$ ($V_{L1}$+$V_{L2}$=0.32 km s$^{-1}$
+0.18 km s$^{-1}$) and the separation speed $\sim$0.43 km s$^{-1}$ ($V_{L3}$+$V_{L2}$=0.25 km s$^{-1}$
+0.18 km s$^{-1}$), respectively.

In panels (b) and (c), the curves of the positive magnetic fluxes within the selected region
indicated by a red rectangle in panel (a) are superimposed on the time-distance diagrams of the loops S1 and S2 in AIA 304 \AA{} channel
to investigate the connections between outflows and magnetic flux changes.
The positive fluxes are calculated, since it can be easily distinguished and their evolutions are more evident.
On the whole, at first there is an increase in the positive fluxes likely due to the fields emerging and combining, then
they decrease. It can be found that there is always a small variation of the fluxes corresponding to occurrence of the strong outflows.
For example, five of them marked by the arrows of $1-5$ in panel (b) are all accompanied by a decreasing of the fluxes.
While for the weak outflows, such corresponding relationship is not so evident. The decrease of positive fluxes should be due
to the negative fluxes invading and cancelling with them.


\section{DISCUSSION AND CONCLUSIONS}
\label{S-Conl}
\begin{figure}
\epsscale{.8} \plotone{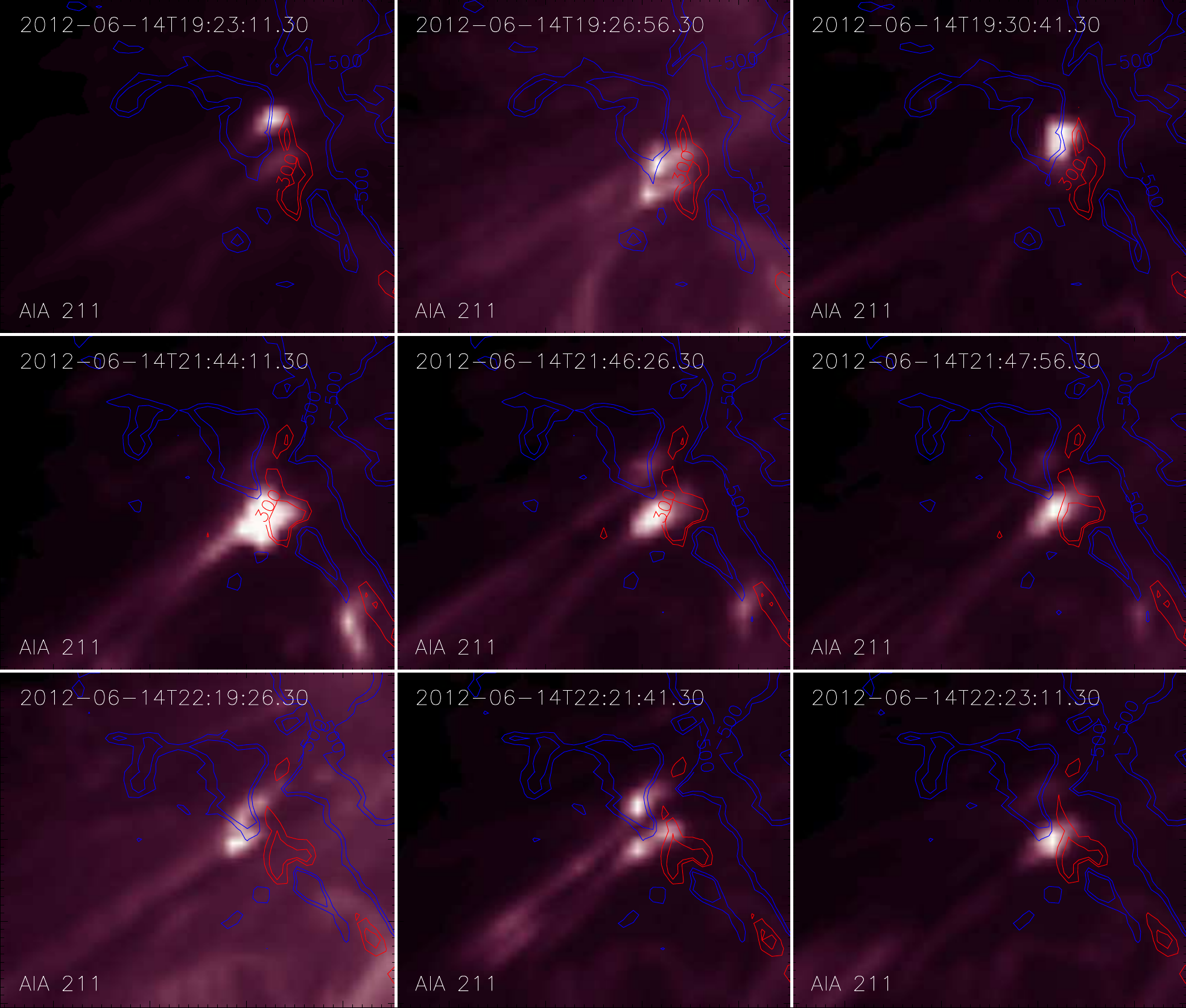} \caption{Images in AIA 211 \AA{} channel at the selected times.
The red/blue contours superimposed on each panel indicate positive/negative HMI LOS magnetic field.}
\label{Fig4}
\end{figure}

\begin{figure}
\epsscale{.8} \plotone{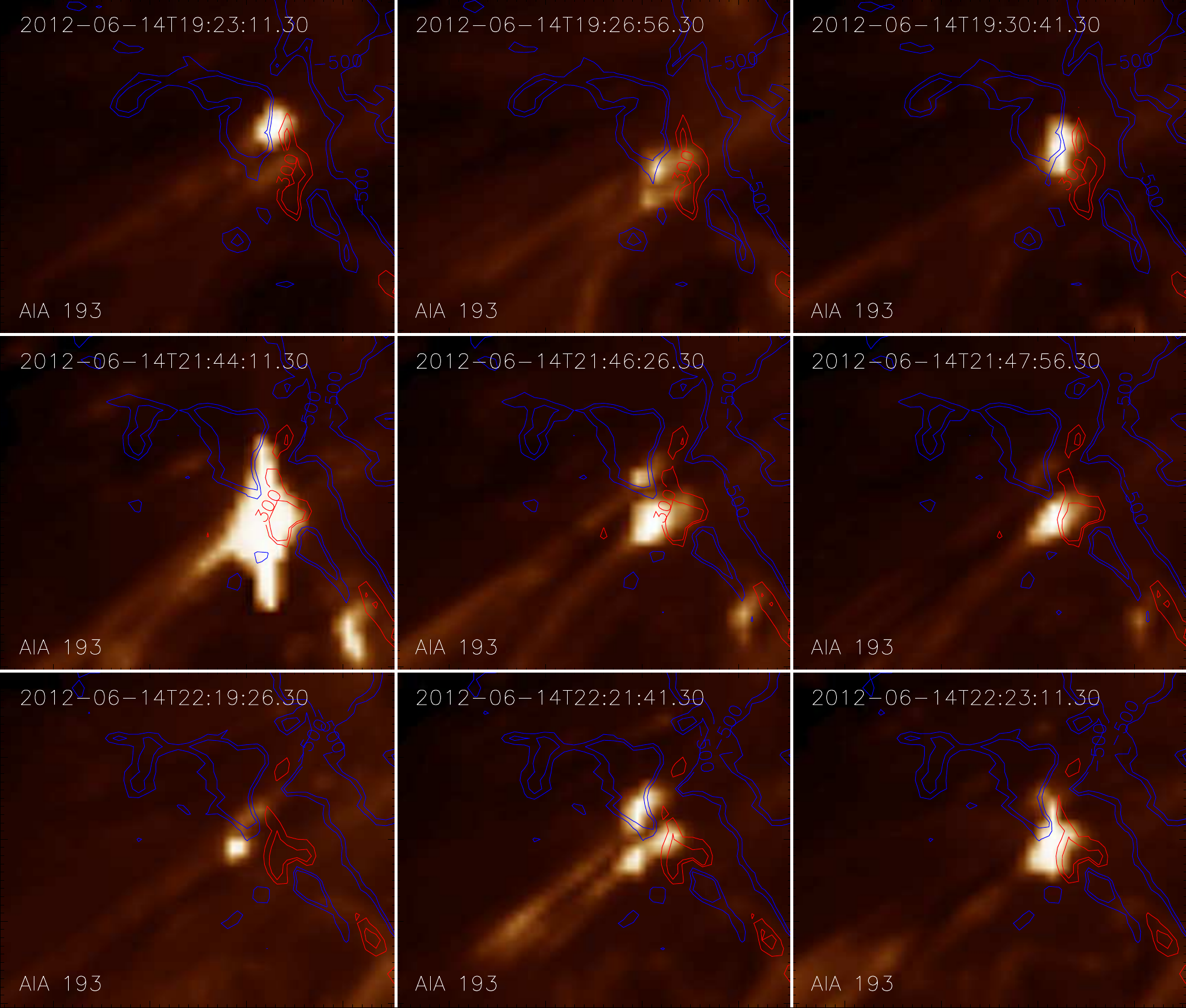} \caption{Same as Fig 4, but for 193 channel.}
\label{Fig5}\end{figure}

\begin{figure}
\epsscale{.8} \plotone{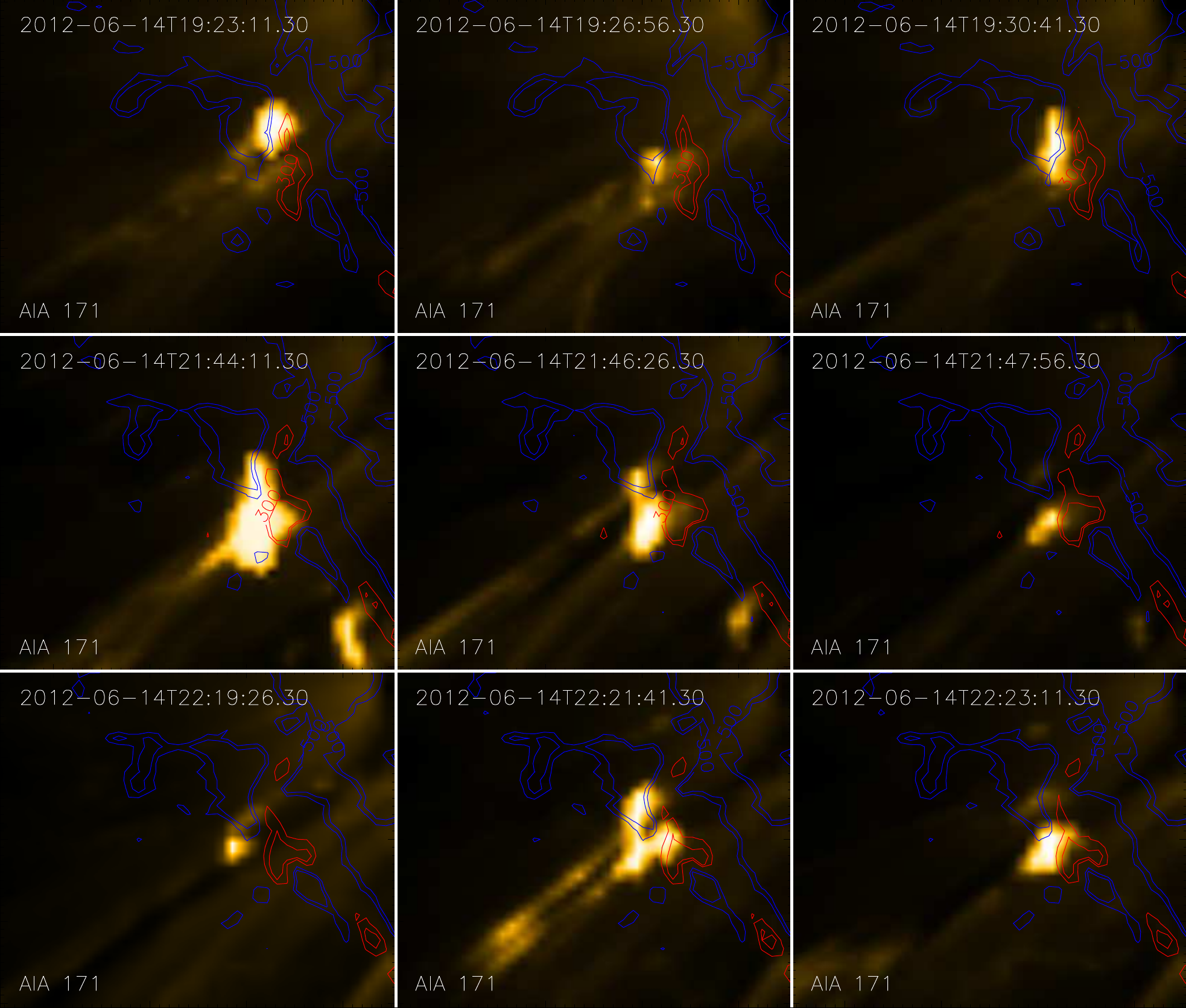} \caption{Same as Fig 4, but for 171 channel.}
\label{Fig6}\end{figure}

\begin{figure}
\epsscale{.8} \plotone{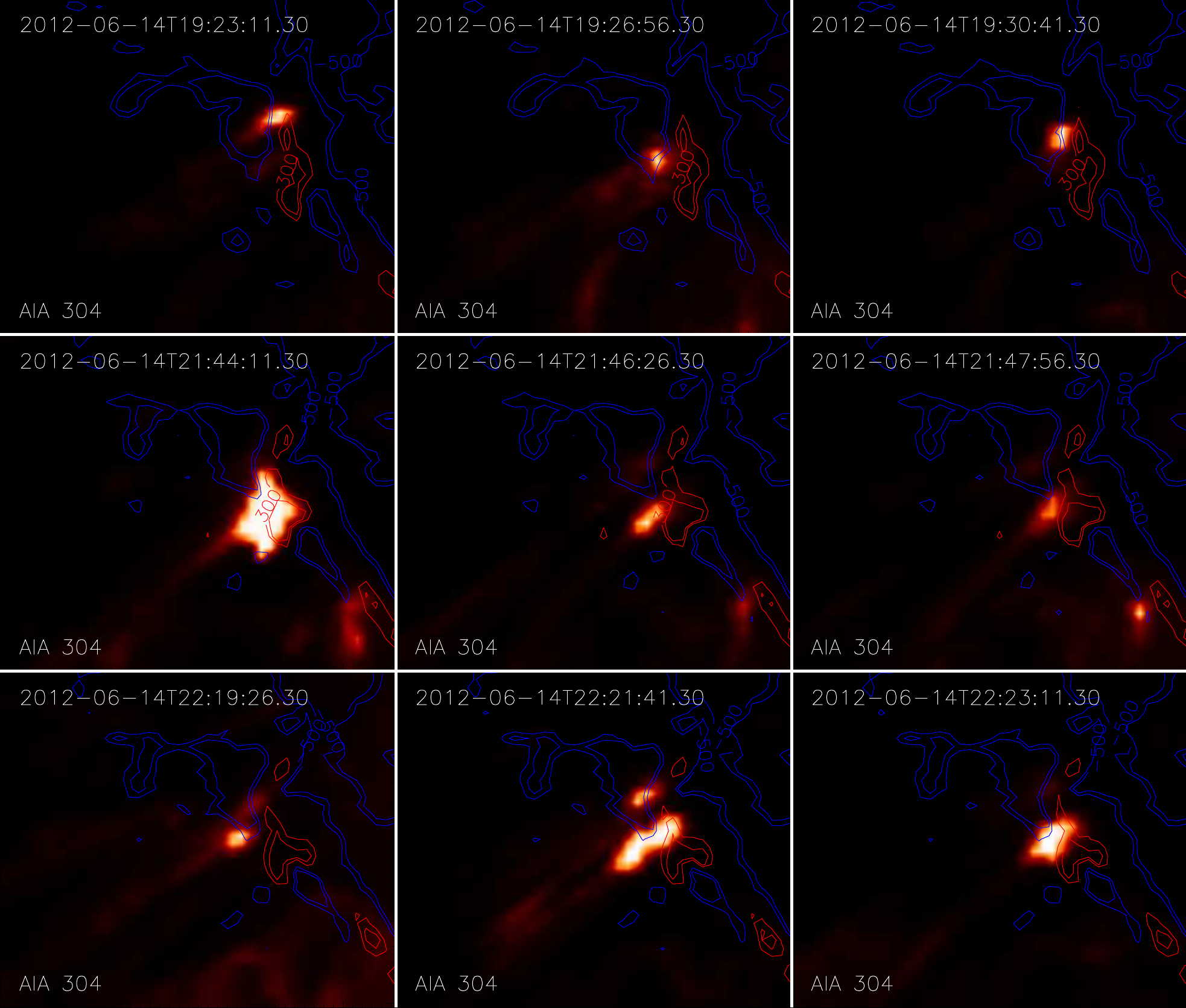} \caption{Same as Fig 4, but for 304 channel.}
\label{Fig7}\end{figure}

\begin{figure}
\epsscale{.8} \plotone{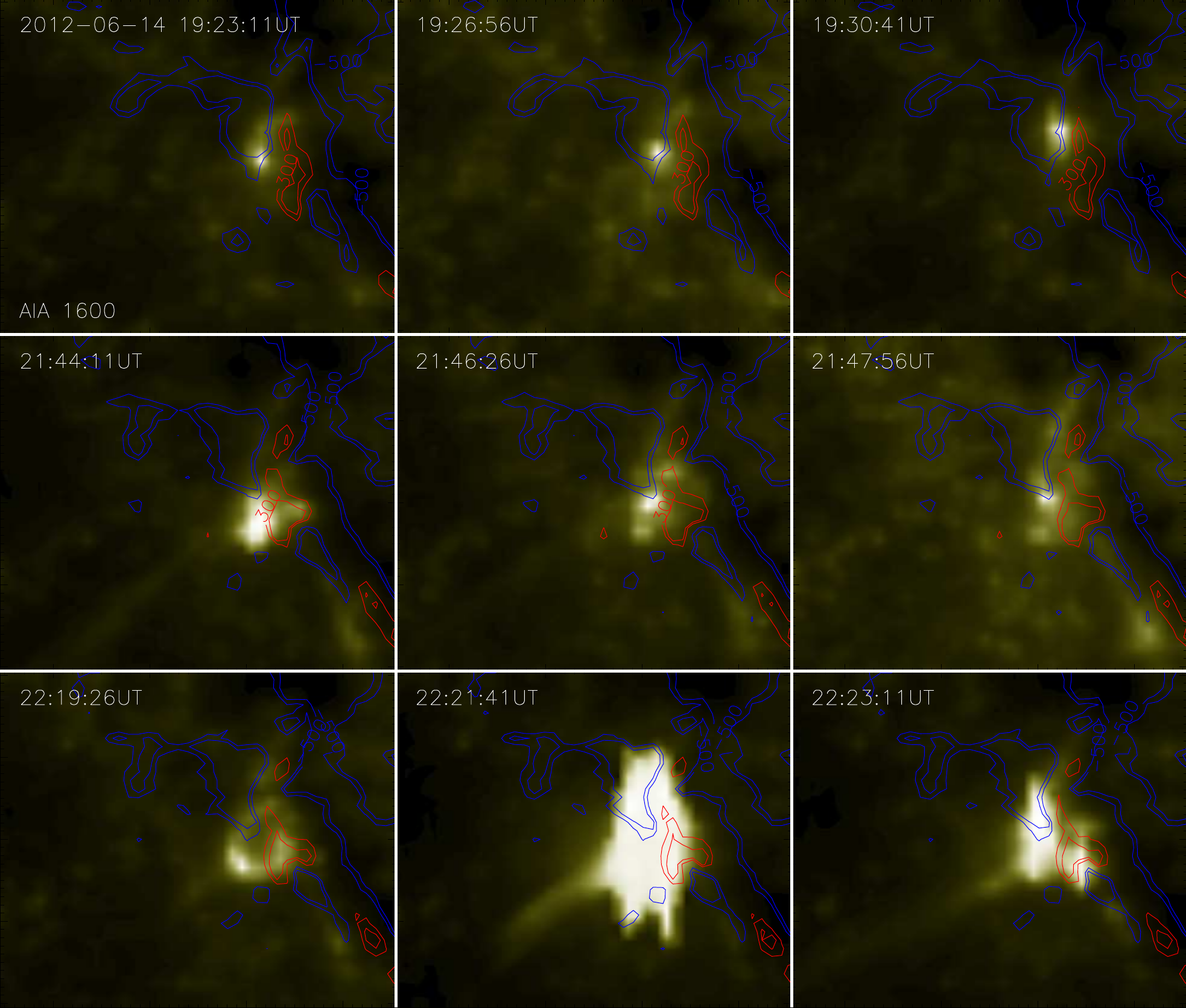} \caption{Same as Fig 4, but for 1600 channel.}
\label{Fig8}\end{figure}

\begin{figure}
\epsscale{.8} \plotone{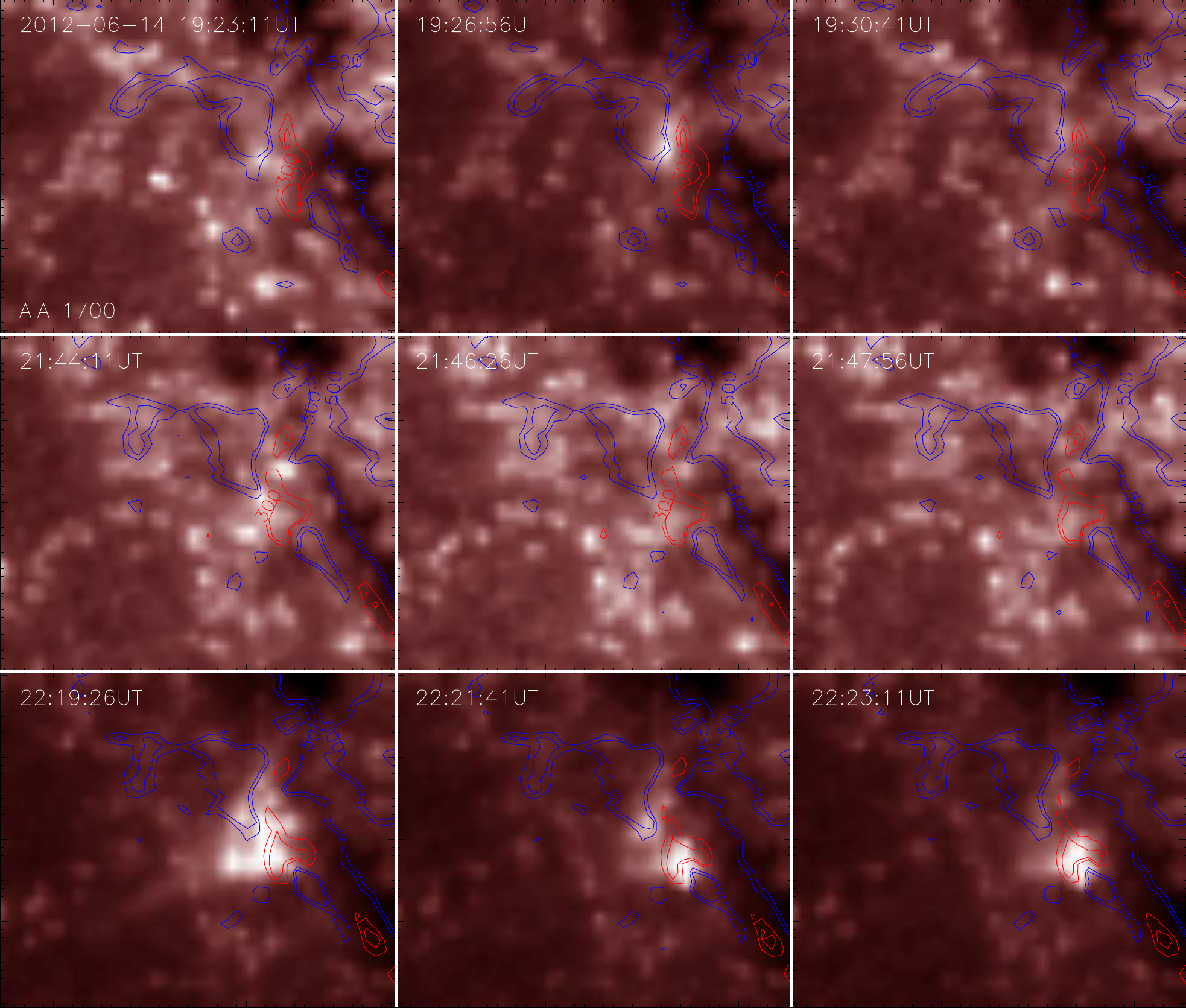} \caption{Same as Fig 4, but for 1700 channel.}
\label{Fig9}\end{figure}

\begin{figure}
\epsscale{1.} \plotone{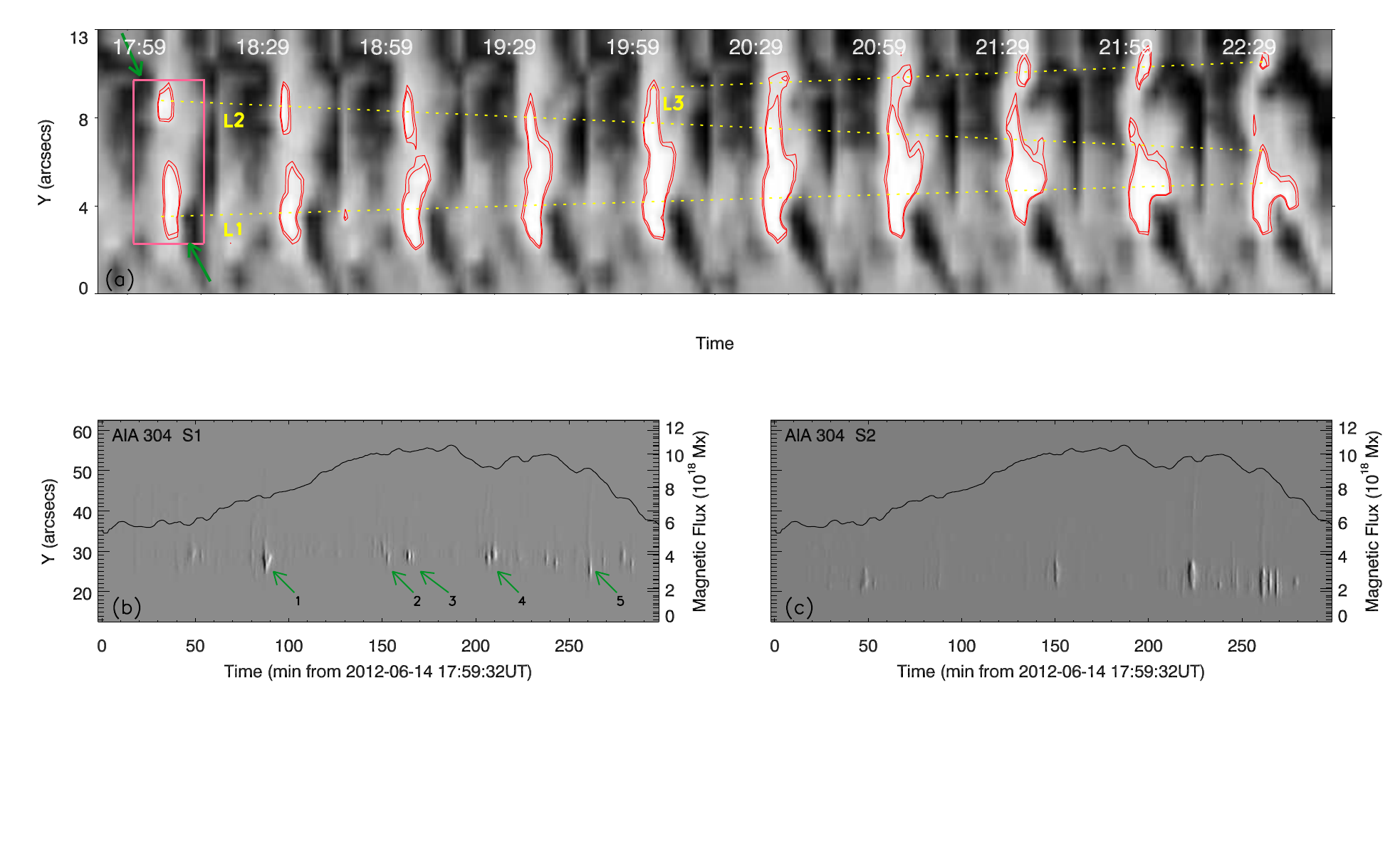} \caption{(a) shows evolutions of the magnetic
field associated with plasma outflows.
(b) and (c) show the evolutions of positive magnetic fluxes within the selected region shown by the red rectangle in panel (a), which are superimposed on the time-distance diagrams in AIA 304 \AA{} channel for the coronal loops s1 and s2 (see Fig 2).}
\label{Fig10}
\end{figure}

In this work with the multi-channel observations by SDO/AIA and HMI, the plasma
outflows and their associated magnetic fields are identified and investigated.
It is found that these plasma outflows can be simultaneously detected in the channels of
211 \AA{}, 193 \AA{}, 171 \AA{} and 304 \AA{}, which means that the outflows are muti-thermal.
By comparing spatial positions of the outflow originating and the mixed magnetic fields
in Fig~\ref{Fig4}-\ref{Fig9}, it can be found that the evident intensity enhancements coincide with
the locations of the magnetic neutral lines. The reconnection positions are likely located at
regions higher than the photosphere as the plasma outflows can not be detected evidently there
(see the data in 1600 \AA{} and 1700 \AA{}). The height of reconnection occurring
may depend on the positions where the two mixed magnetic systems meet and where the necessary
conditions for magnetic reconnections are satisfied. For the current outflows
in this work, the reconnection height should be higher than the photosphere (nearly no signature in 1700 \AA{})
and roughly located at the bottom of the chromopshere (weak signature in 1600 \AA{}), but they can cause a multi-layer atmosphere response as
shown by the simultaneous observations in AIA 211 \AA{}, 193 \AA{}, 171 \AA{} and 304 \AA{} channels.

\begin{figure}
\epsscale{1.} \plotone{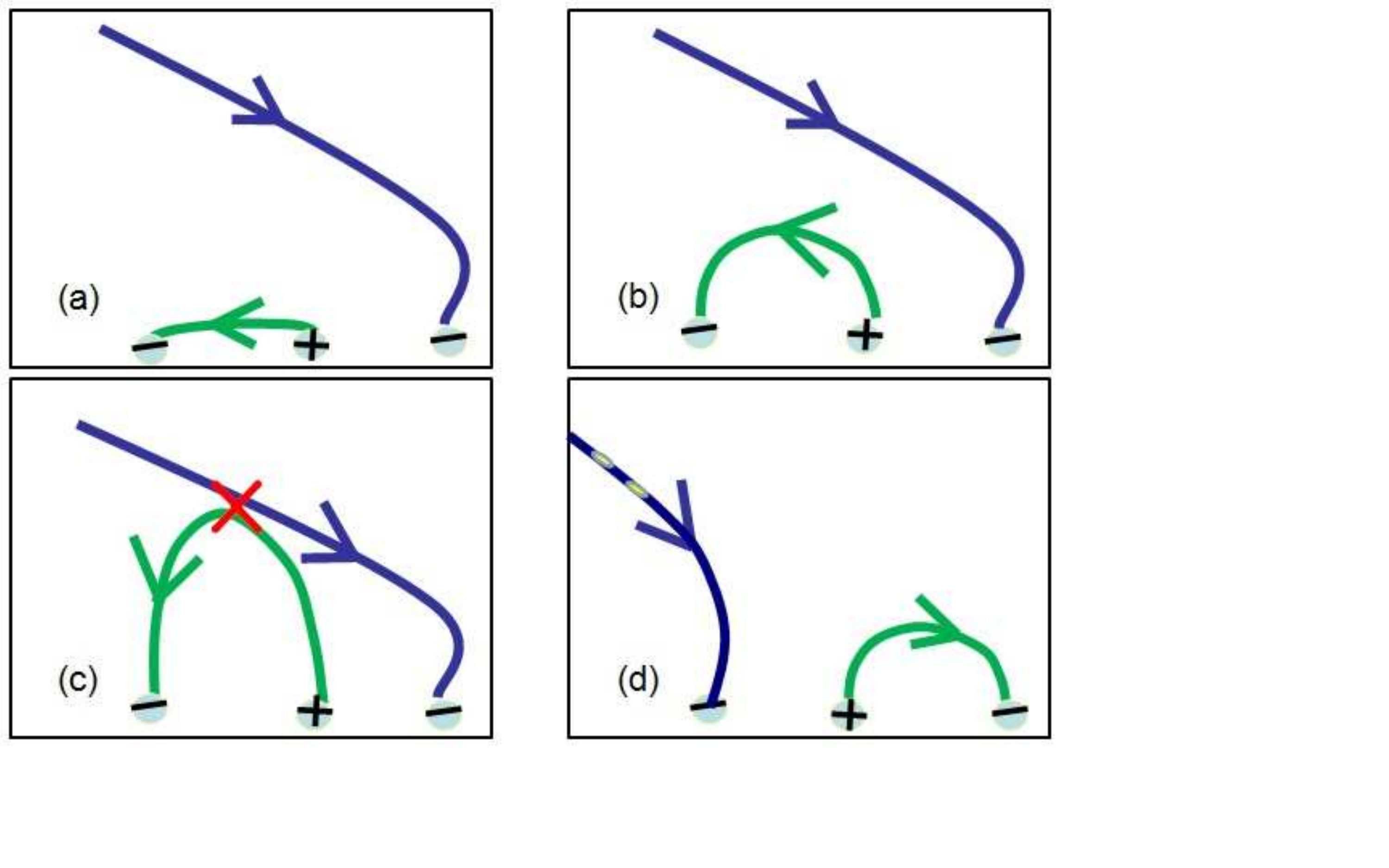} \caption{Cartoon describing magnetic reconnection and the associated outflows.
The green/blue lines represent close/open magnetic filed lines and the polarities are denoted by plus/minus sign. The red cross symbol
indicates the position of magnetic reconnection (c), and the outflows are indicated
by two yellow ellipses along the blue open field line (d).}
\label{Fig11}
\end{figure}

The main conclusions of this work are that the plasma outflows detected in coronal loops
should be the results of magnetic reconnections. The whole event can be explained
simply by some cartoons shown in Fig~\ref{Fig11}. In panel (a), blue lines correspond to the high-altitude open field lines rooted on the edges of main active region, and green lines represent the low-altitude closed field lines connecting two small scale magnetic polarities. Panel (b) shows that the small scale magnetic fields, located close to the edges of the active region, begin to emerge from sub-photosphere. Panel (c) shows that the reconnections occur between these two magnetic systems when the field lines of small scale magnetic fields reach the height
of the large-scale open field lines. Panel (d) displays that outflows are created along the new open field lines
after magnetic reconnection, indicated by two yellow ellipses along the blue open field lines. Here, we take the released magnetic
energy as the driver for the plasmas outflows at Alfven speeds, $100-1000$ km s$^{-1}$ \citep{cir07,nak12}. The presence of these two opposite magnetic systems is of the essential condition for magnetic reconnection. The outflow events can also been interpreted as the interactions between emergence fluxes (green closed lines) with high density material and open fluxes (blue lines) with relative low density material due magnetic pressure gradient, consequently, a bundle of newly opened flux created that allowing significant outflows of escaping plasma \citep{rao08, bak09, har12, su13}.


\acknowledgments This work is supported by the National Basic Research Program of China
under grant 2011CB8114001, the Natural Science Foundation of China (Grant Nos.
11373040,11221063, 11125314, 11173033, 11103037, 11103038, 11203036 and 11178016) and Important Directional
Projects of Chinese Academy of Sciences (Grant No. KLCX2-YW-T04).


\clearpage

\clearpage


\clearpage

\end{document}